\RequirePackage[hyphens]{url}
\documentclass[conference]{IEEEtran}
\usepackage{cite}
\usepackage{amsmath,amssymb,amsfonts}
\usepackage{algorithmic}
\usepackage{graphicx}
\usepackage{textcomp}
\usepackage{svg}
\usepackage{xcolor}
\def\BibTeX{{\rm B\kern-.05em{\sc i\kern-.025em b}\kern-.08em
    T\kern-.1667em\lower.7ex\hbox{E}\kern-.125emX}}

\usepackage{array,booktabs}
\usepackage{pifont}
\usepackage{breakurl}
\usepackage{listings}
\usepackage{soul}
\usepackage{float}
    
\makeatletter
\newcommand{\linebreakand}{%
  \end{@IEEEauthorhalign}
  \hfill\mbox{}\par
  \mbox{}\hfill\begin{@IEEEauthorhalign}
}
\makeatother

\begin{document}
%-------------------------------------------------------------------------------

\date{}
\title{\Large \bf Dead Man's PLC: Towards Viable Cyber Extortion for Operational Technology}

\author{Richard Derbyshire\textsuperscript{a,b}, Benjamin Green\textsuperscript{b}, Charl van der Walt\textsuperscript{a}, David Hutchison\textsuperscript{b} \\ a \{ric.derbyshire, charl.vanderwalt\}@orangecyberdefense.com \\ b \{r.derbyshire1, b.green2, d.hutchison\}@lancaster.ac.uk}

\iffalse
\author{\IEEEauthorblockN{Richard Derbyshire}
\IEEEauthorblockA{\textit{Security Research Center} \\
\textit{Orange Cyberdefense}\\
London, UK \\
ric.derbyshire@orangecyberdefense.com}
\and
\IEEEauthorblockN{Benjamin Green}
\IEEEauthorblockA{\textit{Computing and Communications} \\
\textit{Lancaster University}\\
Lancaster, UK \\
b.green2@lancaster.ac.uk}
\vspace*{1mm}
\linebreakand
\IEEEauthorblockN{Charl van der Walt}
\IEEEauthorblockA{\textit{Security Research Center} \\
\textit{Orange Cyberdefense}\\
London, UK \\
charl.vanderwalt@orangecyberdefense.com}
\and
\IEEEauthorblockN{David Hutchison}
\IEEEauthorblockA{\textit{Computing and Communications} \\
\textit{Lancaster University}\\
Lancaster, UK\\
d.hutchison@lancaster.ac.uk}
}
\fi

\maketitle

%-------------------------------------------------------------------------------
\begin{abstract}
%-------------------------------------------------------------------------------
For decades, operational technology (OT) has enjoyed the luxury of being suitably inaccessible so as to experience directly targeted cyber attacks from only the most advanced and well-resourced adversaries. However, security via obscurity cannot last forever, and indeed a shift is happening whereby less advanced adversaries are showing an appetite for targeting OT. With this shift in adversary demographics, there will likely also be a shift in attack goals, from clandestine process degradation and espionage to overt cyber extortion (Cy-X). The consensus from OT cyber security practitioners suggests that, even if encryption-based Cy-X techniques were launched against OT assets, typical recovery practices designed for engineering processes would provide adequate resilience. In response, this paper introduces Dead Man's PLC (DM-PLC), a pragmatic step towards viable OT Cy-X that acknowledges and weaponises the resilience processes typically encountered. Using only existing functionality, DM-PLC considers an entire environment as the entity under ransom, whereby all assets constantly poll one another to ensure the attack remains untampered, treating any deviations as a detonation trigger akin to a Dead Man's switch. A proof of concept of DM-PLC is implemented and evaluated on an academically peer reviewed and industry validated OT testbed to demonstrate its malicious efficacy.

\end{abstract}

%-------------------------------------------------------------------------------
\section{Introduction}
\label{intro}

%What is OT
There exists a plethora of industry sectors and critical national infrastructure (CNI), from manufacturing to power generation, that use automated monitoring and control of their constituent physical devices via sensors and actuators.
Such capability is achieved through the deployment of operational technology (OT), which senses and manipulates the physical world in real time.
OT is conceptually found between the information technology (IT) that runs the enterprise part of an organisation and the physical devices that comprise its operational part, as defined by the Purdue Enterprise Reference Architecture \cite{Purdue}.

%Current state of OT attacks
Unlike IT, the specialist hardware and software of OT is somewhat more invisible as it is typically restricted to specific environments, including factories and power plants.
For other than site engineers, access to OT assets is usually restricted to simulations or testbeds \cite{Green2020}, notably for cyber security practitioners and adversaries.
The result of this barrier to access is a fortunate dearth of cyber attacks targeting OT when compared with attacks that target IT, even in spite of the manufacturing sector being the predominant target for IT attacks in 2022 \cite{Navigator}.

%What is Cy-X
Looking at IT-targeted cyber attacks in more detail, it is evident that a significant proportion of them use cyber extortion (Cy-X) tactics \cite{Navigator}.
Historically, Cy-X is known for ransomware, which prolifically encrypts data across an IT asset or network of assets, rendering them unusable, and therefore causing a denial of service effect until a ransom is paid for the decryption key.
However, there is a shift towards more creative ways by which cyber criminals extort their victims, hence the encompassing term Cy-X.
The evolution of IT Cy-X tactics includes not only encrypting the victim's data, but exfiltrating the unencrypted data first for the purpose of threatening to leak it, something which is proving effective in the face of the regulatory pressures of data protection \cite{Navigator}.

%OT attacks and the demographic shift
In the rare occurrences of cyber attacks targeting OT, they have often been conducted by either insider threats or state-sponsored adversaries \cite{Derbyshire2018,Miller2021}.
However, 2022 saw an increased \iffalse display of\fi interest in conducting OT-targeted cyber attacks from less typical adversary types, such as \iffalse those that could be categorised as\fi cyber criminals, whose tactical focus is Cy-X \cite{Kaspersky,Fortinet,Forescout}.
Such endeavours will be lucrative to cyber criminals because of the victims' increased willingness to pay to avoid the complexity and scale of the costs associated with OT outages \iffalse, including physical safety, loss of productivity, spoiled product, and environmental damage\fi \cite{Outage}.
Although the delicate and real-time nature of OT means that it faces specific challenges in addressing cyber security risks, typical solutions to engineering issues mean that dedicated OT assets may be resilient to targeted, encryption-based Cy-X attacks like ransomware \cite{Staves2022}.
More specifically, existing practices of replacing a faulty programmable logic controller (PLC) with a new one, and uploading the correct configuration, would likely prove to be effective against encryption based ransomware in the way that regular backups are used to recover from similar IT-targeted ransomware attacks.

%Introduce DM-PLC
This paper introduces Dead Man's PLC (DM-PLC), a pragmatic first step towards a viable Cy-X technique that directly targets OT devices while circumventing the resilience of existing response and recovery tactics.
DM-PLC utilises existing functionality within an OT environment to simultaneously do the following: create a covert monitoring network of PLCs and engineering workstations (EWs) that constantly poll one another; monitor for any deviations from the attack's behaviour; and deny configuration access to the victim.
Should the victim make an attempt to alter the environment under adversary control or not pay their ransom in time, DM-PLC will activate a trigger akin to a Dead Man's switch, causing all PLCs to set their outputs to an ``ON'' state, resulting in chaos within the victim's physical environment.
DM-PLC brings OT-targeted attacks in line with the modus operandi of the emerging demographic of adversaries in the area, i.e. cyber criminals, and reflects the shift from encryption-based Cy-X techniques to more creative ways to conduct ransom attacks.
In doing so, this work highlights the fact that an adversary does not require significant investment, experience, or sophisticated root level access to PLCs to conduct such an OT-targeted cyber attack.
Rather, DM-PLC demonstrates that an adversary may hold an entire OT environment to ransom by simply using existing communications and security features against the victim.

In summary, the main contributions of this work are:

\begin{enumerate}
    \item The proposal of a technique to perform a Cy-X attack on OT devices, utilising only existing functionality, and circumventing resilience encountered in current best practice.
    \item A practical implementation of DM-PLC is demonstrated as a proof of concept on two PLCs and an EW.
    \item DM-PLC is then scaled up and evaluated on an academically peer reviewed and industry validated testbed, identifying its strengths and limitations.
    \item Mitigation techniques are proposed for OT asset owners to protect themselves against such attacks in the future.
\end{enumerate}

The remainder of this paper is structured as follows.
Section \ref{related} chronicles the current state of the art of OT-targeted Cy-X capability.
Section \ref{background} establishes an understanding of the OT under attack and the preconditions for the attack to take place.
Section \ref{approach} provides a discussion of the conceptual approach to DM-PLC.
Section \ref{implementation} describes a proof of concept implementation on two PLCs and an EW.
Section \ref{evaluation} presents the evaluation, its results, and ensuing discussion.
Section \ref{mitigation} proposes the types of controls that may be utilised to mitigate attacks such as DM-PLC.
Finally, Section \ref{conclusion} summarises what has been proposed and reflects on the DM-PLC technique in a concluding discussion.

%\vspace*{-3mm}
\section{Related Work}
\label{related}
%\vspace*{-3mm}

Historically, OT cyber attacks have seldom involved the deployment of malware specifically targeting OT devices, such as PLCs, remote terminal units (RTUs), or human machine interfaces (HMIs) \cite{Derbyshire2018,Miller2021}, with the notable exceptions of highly complex attacks such as Stuxnet \cite{Langner2011}.
More recently, OT-specific malware has been discovered, having been deployed to facilitate other advanced attacks, including examples such as CRASHOVERRIDE \cite{DragosCrash}, TRITON \cite{NCSCTriton}, Industroyer2 \cite{ForescoutIndustroyer2}, and PIPEDREAM \cite{DragosPipe}.
However, these attacks do not fit the modus operandi of cyber criminals looking to enter the space, who typically pursue financially motivated engagements.
Moreover, it is well documented that such attacks would have required a level of process comprehension \cite{Green2017,Green2021} that would be considered inaccessible to this new, incoming demographic of adversaries.

When identifying novel tactics, techniques, and procedures (TTPs) for targeting OT, or the malware used, academic literature is sparse - particularly for examples that are both pragmatic and suitable to cyber criminals looking to profit from their engagements without requiring costly process comprehension.

Formby et al. \cite{Formby2017} introduce LogicLocker, stating that it is ``The first known example of ransomware to target PLCs in industrial control system networks''.
LogicLocker's approach involves exploiting vulnerabilities in PLCs discovered on Shodan, moving laterally and horizontally within the OT network, locking the affected PLCs, and encrypting the PLCs' configurations, before deploying a `logic bomb'.
Some practical concerns about this approach include the reliability of discerning PLCs on Shodan (many are honeypots \cite{Ercolani2016, Maesschalck2021}), the noise of using these exploits against the PLCs (particularly within the OT environment), and how some of the internal scanning and lateral movement techniques are achieved without root access to the PLCs.

Not dissimilar to LogicLocker is ICS-BROCK, introduced by Zhang et al. \cite{Zhang2020}, which is intended to be stealthy and practical for application in any environment, with any type of PLC.
ICS-BROCK's approach follows a similar set of steps to LogicLocker, whereby the malware detects and exploits Windows-based vulnerabilities in the OT network, identifies PLCs via the address resolution protocol (ARP), locks the affected PLCs, deploys a `logic bomb', and encrypts the engineering workstation.
As with LogicLocker, ICS-BROCK will cause significant noise by attempting to exploit common vulnerabilities (MS17-010, MS08-067, CVE-2019-0708).

LogicLocker and ICS-BROCK are commendable contributions to the concept of OT-specific Cy-X, but they also leave room for improvement before being considered truly pragmatic in a live engagement. 
Both put considerable effort into justifying the malware's access to the OT environment. 
However, there is no shortage of access in modern attacks \cite{Kaspersky,Fortinet,Forescout}, the real challenge to overcome being the creative use of the OT assets themselves.
Both methods purport to exploit commonly known vulnerabilities early on in their approaches, before the payloads have been delivered, which increases the risk of detection prior to the point of extortion.
Along with the above assumptions, neither approach considers the non-trivial amount of process comprehension required to ensure the success of even a simple OT-targeted attack \cite{Green2017,Green2021}.
Finally, neither approach considers the prospect of response and recovery processes in an operational environment.
Zhang et al. \cite{Zhang2020} state ``Something we need to emphasize is that, once the PLC is locked, the operator will not choose to re-flash the PLC for it would make ICS stop, stop always means big economic losses.'' - in fact, this is \textit{exactly} what the operator would do \cite{Staves2022}.

DM-PLC, this paper's novel approach for conducting Cy-X against OT assets, learns from LogicLocker and ICS-BROCK, while iterating on the weaknesses discussed above by focusing on employing only existing and expected OT functionality and further using it to counter response and recovery practices typically encountered in an OT environment.

%\vspace*{-3mm}
\section{Threat Model}
\label{background}
%\vspace*{-3mm}

To effectively introduce DM-PLC, it is first necessary to establish an understanding of the OT under attack as well as the preconditions required for the attack to take place.

\subsection{DM-PLC OT attack surface}

DM-PLC is focused on the existing intercommunication between OT assets, predominantly engineering workstations (EWs) and PLCs. 
However, it could be extended to additional OT assets such as remote terminal units (RTUs) and human machine interfaces (HMIs) should the adversary so choose.

From a technical perspective, EWs are generally Windows-based devices complete with the typical vulnerabilities encountered in enterprise IT assets.
The key aspect about an EW is that it is enhanced with industrial configuration software, which allows it to communicate via industrial protocols and configure embedded OT devices such as PLCs.
For example, in a Siemens ecosystem, this industrial configuration software would be the Totally Integrated Automation (TIA) Portal \cite{TIA}.

PLCs are embedded devices that can have a variety of architectures.
These devices typically have minimal operating systems, such as modified versions of Linux or even bespoke implementations, that are inaccessible to both the operators and potential adversaries.
However, at the level which is accessible to the operator, there is a wealth of functionality available such as ICMP, SNMP servers, and HTTPS servers.

PLCs are programmed by the industrial configuration software on an EW, using the industrial protocols that are enabled by the software.
The program, pushed to the PLC by the EW, controls what the PLC does and how it senses and controls the physical process.
It can be written in a variety of languages such as ladder logic and can be entirely bespoke; however, default library functions provided by the PLC vendor are commonly used \cite{Green2021}.

\begin{figure}[h!]
    \centering
    \includegraphics[width=0.85\linewidth]{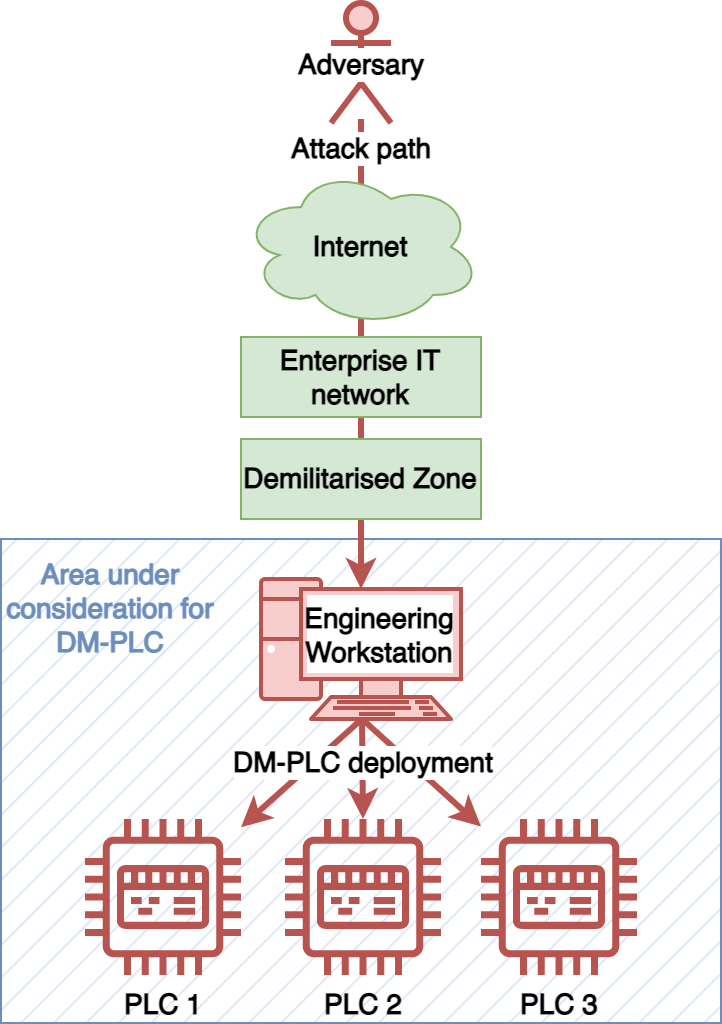}
    %\vspace*{-1mm}
    \caption{DM-PLC attack scenario}
    %\vspace*{-2mm}
    \label{fig:scenario}
\end{figure}

\subsection{DM-PLC attack scenario}

Modern adversaries do not appear to be encountering challenges when navigating to the OT environments of their victims \cite{Kaspersky,Fortinet,Forescout}, and no value or novelty can be derived from describing that process.
Therefore, this paper does not concern itself with initial access or lateral movement prior to deploying DM-PLC. 
It is instead assumed that the adversary has completed these tactics and is already in position.
As such, this is reflected in Figure \ref{fig:scenario} where the adversary's traversal to the OT environment is depicted in a simplified manner.

The physical and logical layout of an OT environment can vary significantly between organisations, and this variation can be further exacerbated between sectors \cite{800-82}; for that reason, Figure \ref{fig:scenario} depicts a high level attack scenario complete with the necessary trajectory required for the adversary, up to and including the deployment of DM-PLC.
To begin the deployment, the adversary must have access to an EW in the OT environment.
From here, they will typically have access to reconfigure a fleet of PLCs controlling at least one process.

%\vspace*{-3mm}
\section{The DM-PLC Approach}
\label{approach}
%\vspace*{-3mm}
While operating under the assumed compromise of an EW provides a valuable starting point, the DM-PLC approach must remain flexible to allow for context and vendor agnostic development. 
EWs typically contain a wealth of information, and more specifically a detailed view of all operational PLCs and their associated configuration~\cite{Green2017}. 
This section describes how such information affords a sufficient level of process comprehension to develop bespoke DM-PLC deployments, tailored to the system under attack. 
Moreover, as EWs offer direct connectivity to each PLC, they present a trusted conduit through which DM-PLC can be deployed and executed. 

Taking existing work as both inspiration and lessons to be learned, DM-PLC seeks to further develop the existing notion of OT Cy-X through the employment of legitimate, vendor provided PLC functionality.
The key feature of DM-PLC is circumventing the current reslience of best practice response and recovery, such that removing affected PLCs is not an option without consequence for the victim.
This also means that DM-PLC must not negatively impact the operational process unless it is tampered with or its ransom timer has expired; immediately experiencing negative consequences may dissuade the victim from paying in lieu of replacing the affected devices.
Therefore, the following outlines a target set of high-level DM-PLC requirements:

\begin{enumerate}
    \item Deployable with minimal pre-requisites from an EW.
    \item Runs in parallel to existing operational code.
    \item Does not impact existing operational code.
    \item Is resilient to tampering/response and recovery processes.
    \item Includes tamper detection.
    \item Can enact undesirable wide-spread operational impact.
    \item Requires a key to relinquish control back to system owners.
    \item Can be tested prior to being armed.
\end{enumerate}

The following subsections adopt these requirements as a baseline, and provide details on the proposal of DM-PLC from a conceptual perspective. The individual phases are grouped into preparing, deploying, and arming DM-PLC.

\subsection{Preparing DM-PLC}
The first group of phases for DM-PLC involve conducting necessary reconnaissance and enumeration to build up the minimum required process comprehension to conduct the attack.

\subsubsection{Identify and Validate the Current PLC Project}
\label{a}
As DM-PLC is designed as an extension to existing, trusted PLC code, it is important to first locate the current live codebase, and obtain a better understanding of the overall network architecture (i.e. how many PLCs there are within the environment, with their latest configuration). 
To do this, PLC vendors typically allow for the collation of multiple PLC configuration objects within the same project \cite{FactoryTalk}. 
Consolidation of all PLC code provides engineers with a single reference point when diagnosing issues or extending/enhancing operational functionality. 
This diagnostic capability is of particular importance when validating project files, allowing an adversary to compare ``online'' live PLC code with ``offline'' project files, prior to the addition of DM-PLC to each PLCs codebase \cite{OnlineOffline}. 
Through the use of this feature, the adversary is able to validate the project file and also the EW's ability to connect directly with each operational PLC.

The phase described above gathers a significant amount of the required information for DM-PLC's process comprehension, using only the EW.
The process comprehension itself allows the attack to be further crafted such that it does not interfere with existing operational code, while running alongside it, and when needed makes the biggest impact it can given its circumstances.
Therefore, this phase contributes to satisfying requirements 1, 2, 3, and 6 via preparatory reconnaissance and enumeration.

%1 2 3 6 

\subsubsection{Identify Pre-Existing PLC-PLC Relationships}
\label{b}
Section~\ref{a} established the EW's ability to communicate with operational PLCs via online/offline diagnostic tooling. 
However, this does not mean each and every PLC will be able to communicate with one another, which is a key requirement to increase the resilience/effectiveness of DM-PLC. 
Fortunately, as with the online/offline features, vendors have this covered with device and network views \cite{PROFINETS7}. 
These features provide a network schematic of all PLCs within the project. 
Reviewing the network schematic, individual device IP address details, and identifying the existence of communication libraries \cite{PUTGET} within each PLCs codebase, will allow the adversary to understand existing PLC-PLC relationships, and where new ones could be formed.

Through the completion of this phase, the adversary can enhance their level of process comprehension on PLC-PLC communications, a critical requirement when building resilience and tamper proofing into the DM-PLC attack.
Furthermore, this is again conducted using only the EW, meaning it contributes to satisfying requirements 1, 4, 5, and 6.

%1 4 5 6 

\subsubsection{Identify Core Code Blocks}
\label{c}
PLC code operates in cycles, and these cycles need to be understood. 
Depending on the vendor, PLC programmer, and operational requirements, these will differ both in the terminology used and their function. 
Taking the Siemens ecosystem as an example, OB1 (Organisation Block 1) is the main code block that is being cyclically executed at all times. 
However, OB1 can be interrupted by other code blocks, e.g. OB30 (Cyclic Interrupt). OB30 can interrupt OB1 at a regular time interval to execute a separate block of code. 
It is important for the adversary to understand these code blocks and their sequencing before deploying DM-PLC, thus ensuring that DM-PLC is able to operate as expected in armed mode (i.e. waiting for the victim to pay, and ensuring that they are not enacting response and recovery processes) and triggered mode (i.e. should the victim fail to pay the ransom, resulting in undesirable operational impact) \cite{Programming}.

With the identification of code blocks, the adversary completes the final stages of process comprehension necessary to begin deploying DM-PLC in harmony with the existing PLC code, while ready to interrupt its execution, and cause undesirable operational process disruption should it be tampered with, or exceed the specified time limit.
This phase, therefore, contributes to satisfying requirements 1-6.

%1 2 3 4 5 6 

\subsection{Deploying DM-PLC}
In the second group of phases, the functionality of DM-PLC is built and subsequently deployed to the target devices.

\subsubsection{Introduce PLC-PLC Communications}
\label{d}
When holding an entire OT network to ransom, it is important to identify any attempt from the victim to regain control. 
To do this, DM-PLC requires a covert monitoring network in which all devices communicate with each other to ensure their integrity remains intact and under the control of the adversary. 
In Section~\ref{b} pre-existing relationships were identified, which provided a view on the use of vendor provided communications library functions enabling PLC-PLC communications. 
These same functions are used here to establish new PLC-PLC sessions, allowing for the exchange of status data, providing holistic visibility of all operational devices. 
Where no existing PLC-PLC communications exist, vendor provided functions would still be used. 
However, should one library function fail due to firewall restrictions, for example, re-configuration using an alternative may be required. 
Vendors typically provide a range of communication functions within their libraries, affording adversaries with multiple options and avoiding the need for custom code development \cite{PUTGET,TSEND}.

Unlike prior OT-targeted Cy-X techniques that intend to disrupt PLCs individually \cite{Formby2017,Zhang2020}, DM-PLC focuses on treating the entire process as the entity under ransom.
Therefore this phase is crucial to DM-PLC, acting as the heart of its resilience against tampering or any form of response and recovery, and laying the foundations for satisfying requirements 4 and 5.

%4 5 

\subsubsection{Introduce Engineering Workstation-PLC Communications}
As with PLC-PLC communications, DM-PLC's covert monitoring network would not be complete without the EW. 
Section~\ref{a} validated the connectivity an EW has with each operational PLC; here DM-PLC leverages this connectivity to provide each PLC with a view of the EW's state, and vice versa. 
Each PLC vendor will choose a specific network protocol for EW-PLC configuration management, and as this network protocol is permitted through existing network-based controls (e.g. a firewall), it should be used to establish covert EW-PLC communications. 
Fortunately, there exists a broad range of open source communications libraries that can be used for this purpose with minimal effort \cite{Green2020}. 
The selected library must be installed on the EW. Fortunately they are often very lightweight and will not impact EW performance.

This phase is similar to Section \ref{d} in that it is crucial to DM-PLC's resilience to tampering or response and recovery attempts.
Therefore, this phase also contributes to satisfying requirements 4 and 5. 

%4 5

\subsubsection{Introduce Device Status Checkers}
\label{f}
Once PLC-PLC and EW-PLC communications have been established, a polling and status checker system must be implemented. 
Like any software, PLCs allocate memory to store variable states. 
Here each PLC requires a block of memory to store neighbouring device states and an alert state. 
Starting with the EW, it is required to send a poll to each PLC every second. 
To keep things simple, this could be a binary state that changes from a 0 to a 1 and back again. 
This is written to a designated area of memory in each PLC. 
Each PLC monitors this area of memory for state changes, and if it fails to see a state change it knows something unexpected has happened to the EW (e.g. the victim has disconnected it from the network while enacting response and recovery processes). 
Each PLC must also be configured to establish a polling system such as this with its neighbouring PLCs. 
An additional alert state is also required, especially for scenarios in which some PLCs are isolated from others. 
For example, if there are three PLCs, and PLC 1 is unable to see PLC 3 directly, it will rely on PLC 2 or the EW. 
Should PLC 2 fail to send a poll to PLC 3 it will change its alert state to indicate something has gone wrong. 
Again, to keep things simple, this could be a binary state where 0 is normal, and 1 is alert. 
PLC 1 would not only be actively polling PLC 2, providing it with its current state, but it would also be monitoring PLC 2's status bit, and upon seeing it change to a 1, would immediately stop any of its own polling, and set its alert state to 1. 
This would quickly propagate through the covert communications network, with all devices being made aware of the unexpected change in the environment.

This phase makes use of the covert PLC-PLC and EW-PLC monitoring networks that have been created to ensure that every asset under ransom has not been tampered with.
Therefore, this phase combines with the previous two to satisfy requirements 4 and 5.

%4 5

\subsubsection{Introduce Code Supporting Operational Process Disruption}
Section~\ref{c} identified core code blocks within each PLC. 
There are multiple ways in which these could be manipulated based on the level of process comprehension an adversary has~\cite{Green2017}. 
A basic approach could be to disable its execution and switch all PLC outputs to an ``ON'' state. 
Introducing a normally closed contact before each core code block would only permit its execution when the alert is in a 0 (normal) state, and conversely prevent its execution upon switching to a 1 (alert) state. 
A malicious code block is then required to operate in an opposing manner (i.e. a normally open contact based on the alert state would be introduced as a trigger). 
A review of each PLC's hardware profile within the project file is required to build a malicious code block, through which identification of all output cards and their associated addressing is made possible. 
The malicious code in this example would simply use these addresses to manipulate physical outputs, and in turn, the operational devices they are connected to (e.g. motors, conveyor belts, and valves). 
In addition to the alert state being triggered by the victim's attempts to regain control of their devices, it will also be triggered by a timer. 
The timer duration is set to match the period of time within which the adversary has given the victim to make a payment.

This phase provides DM-PLC with the capability to cause undesirable operational impact should the attack be tampered with or the ransom not be paid in time, which is absolutely necessary for DM-PLC to be a credible Cy-X threat to its victims.
Therefore, this phase satisfies requirement 6.

%6 

\subsubsection{Prevent Victims from Reversing all Changes}
\label{h}
Continuing the approach taken to the development of DM-PLC thus far, preventing the victim from reverting the changes applied to their devices should make use of vendor provided functionality - essentially turning the victim's decision not to use these features against them. 
For example, PLCs are increasingly being provided with password protection capabilities, something taken advantage of by Zhang et al. \cite{Zhang2020}. 
In addition, PLC project files also offer a variety of protection features to password protect and encrypt their contents \cite{SiemensSecurity}. 
Finally, the EW should lock out all trusted users and encrypt its disk using conventional ransomware techniques \cite{Oz2022}. 
These, in conjunction with DM-PLC's covert monitoring network, make response and recovery extremely difficult without experiencing undesirable operational impact.

Restricting the victim's access to the PLCs and EWs that have been affected during the attack is the final preparatory phase of DM-PLC.
This provides an additional measure to prevent any tampering or response and recovery processes, on top of the covert PLC-PLC and EW-PLC monitoring networks.
Not only does the password protection on the PLCs and project files add another layer to requirement 4, but they work in conjunction with the traditional encryption-based ransomware techniques used on the EW to provide a mechanism for satisfying requirement 7.

%4 7
\subsection{Arming DM-PLC}
Once DM-PLC's functionality has been deployed to the affected devices, the final phase is to arm its Dead Man's switches.

\subsubsection{Arm DM-PLC}
A staged approach to the deployment of DM-PLC is required to ensure it operates as expected. 
Deploying PLC and EW code in a structured way, as described across Sections~\ref{d} to~\ref{h}, provides a starting point. 
In addition to this, a normally open contact mapped to an ``Enable'' variable is required before all new PLC code blocks (i.e. communication and malicious code). 
This will prevent any alerts from being raised until it is set to a 1 (enable) state, which can be done across all devices simultaneously by the EW once the adversary is ready. 
Collectively, this strategy of gradual code deployment, testing, and enabling will put DM-PLC in an armed state, with the alert state being a trigger to cause undesirable operational process disruption. 
Upon receipt of payment, the victim will be provided with a key that unlocks the EW, allowing DM-PLC to be disabled via the enable variable in each PLC, and removes all password protections from their PLCs and their PLC project files.

The final phase of DM-PLC is focused on testing and safe deployment such that it is not triggered prematurely during the setup phases.
The enable variable doubles up as an easy disarm mechanism once the adversary has relinquished control of the OT environment on full receipt of payment.
This phase, therefore, assists the one discussed in Section \ref{h} that satisfies requirement 7, along with fulfilling requirement 8.

%7 8
%\vspace*{-3mm}
\section{Implementation}
\label{implementation}
%\vspace*{-3mm}
Section~\ref{approach} introduced the concept of DM-PLC, defining key characteristics and how they map onto a set of baseline requirements. This section builds on DM-PLC as a conceptual construct and demonstrates how it can be deployed in practice on a real-world system.

To best describe a practical implementation of DM-PLC and avoid unnecessary complexity, the minimum infrastructure required to conduct the attack is used; just two PLCs and one EW. The PLCs are a Siemens ET200S and a S7-1200, and the EW is running the Siemens TIA Portal V17 programming agent \cite{TIA}.

To provide consistency with Section~\ref{approach}, the following subsections provide a mirrored structure, populated with practical, applied content. Figures depicting the implementation process are referred to throughout this section for additional clarity.

\subsection{Preparing DM-PLC}

\subsubsection{Identify and Validate the Current PLC Project}
\begin{figure}[H]
    \begin{center}
    \includegraphics[width=0.5\linewidth]{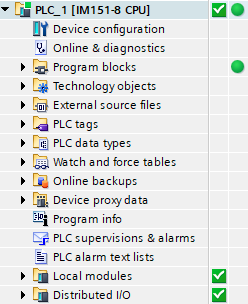}
    \end{center}
    \vspace*{-4mm}
    \caption{Online Mode}
    \vspace*{-3mm}
    \label{fig:online}
\end{figure}

Upon loading TIA Portal, the adversary will be presented with a list of recently opened projects. If no projects are listed, a basic search of local and remote drives for `.ap17' can be performed. By default, the TIA Portal GUI is split into three sections, with the first section displaying a tree structure of all associated devices and their configurations. In order to validate the applicability of a project with each PLC the ``Online'' function can be used. By right clicking each PLC and selecting ``Go Online'', TIA Portal will compare the project codebase with that of the PLC, with a set of green indicators confirming a 100\% match (see Figure~\ref{fig:online}).

\subsubsection{Identify pre-existing PLC-PLC Relationships}
\begin{figure}[H]
    \begin{center}
    \includegraphics[width=0.7\linewidth]{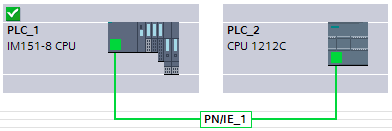}
    \end{center}
    \vspace*{-4mm}
    \caption{PLC-PLC Relationships}
    \vspace*{-3mm}
    \label{fig:relationships}
\end{figure}

The ``Devices and Networks'' feature in TIA Portal can be used as a quick reference point in the determination of existing PLC-PLC relationships (see Figure~\ref{fig:relationships} depicting PLC\_1 and PLC\_2 on the same network). A review of each PLCs IP addressing and core code base can also be undertaken for further validation, and to identify which communications protocols are in use. An example of this would be the PUT communication library function provided by Siemens within TIA Portal (see Figure~\ref{fig:put}). The identification of a function such as this between two PLCs, confirms the use of S7-Comm as a permitted network protocol.

\subsubsection{Identify Core Code Blocks}
\label{ccb}
\begin{figure}[h!]
    \begin{center}
    \includegraphics[width=0.5\linewidth]{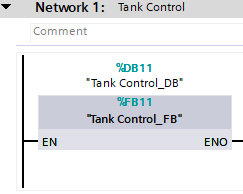}
    \end{center}
    \vspace*{-4mm}
    \caption{Core Code Block}
    \vspace*{-3mm}
    \label{fig:ob1}
\end{figure}

To keep this proof of concept (PoC) simple, there is only one core code block (a tank control Function Block) on PLC\_1, residing in OB1, the main program cycle in Siemens PLCs, as described in Section~\ref{c} (see Figure~\ref{fig:ob1}). There are no pre-requisites to this code block's execution (nothing to the left of the Function Block in Figure~\ref{fig:ob1}), meaning it will execute on every cycle of the PLC's codebase. There exist no additional code blocks that can interrupt OB1's cycle on this PLC. PLC\_2 contains the same codebase, with no additional interrupts or functions to consider.

\subsection{Deploying DM-PLC}

\subsubsection{Introduce PLC-PLC Communications}
The first step in developing DM-PLC is to establish PLC-PLC communications. To do this an appropriate communications function must be installed on each PLC. For this PoC, the TIA Portal-provided communications library functions PUT and GET are used.

\begin{figure}
    \begin{center}
    \includegraphics[width=0.8\linewidth]{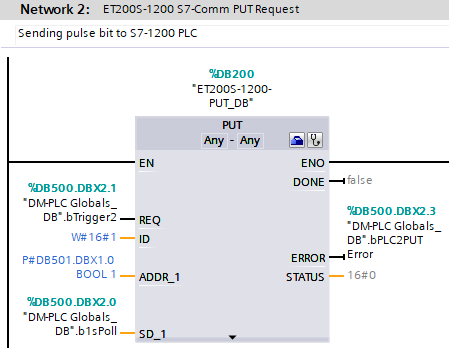}
    \end{center}
    \vspace*{-4mm}
    \caption{S7-Comm PUT Setup}
    \vspace*{-3mm}
    \label{fig:put}
\end{figure}

\begin{figure}
    \begin{center}
    \includegraphics[width=0.8\linewidth]{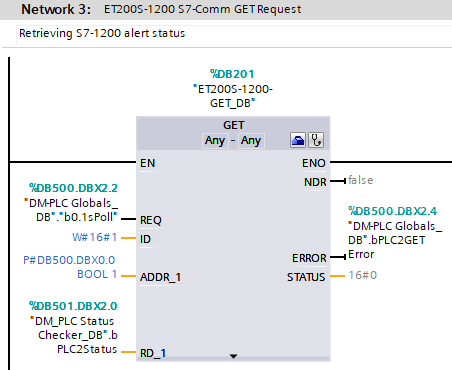}
    \end{center}
    \vspace*{-4mm}
    \caption{S7-Comm GET Setup}
    \vspace*{-3mm}
    \label{fig:get}
\end{figure}

The PUT library function is setup on PLC\_1 to send data to PLC\_2 (see Figure~\ref{fig:put}). The GET library function is setup on PLC\_1 to retrieve data from PLC\_2 (see Figure~\ref{fig:get}). To ensure communications are flowing between the two devices, testing is required at this stage. Setting up dummy data to send and request gives an adversary confidence in the deployment of any given communications library, prior to its use as a primary method of data exchange within the covert DM-PLC network.

\subsubsection{Introduce Engineering Workstation-PLC Communications}
\label{EWPLCComm}
Once the PUT/GET library functions have been deployed, the EW has proven it is able to communicate via the S7-Comm protocol with each PLC (the protocol TIA Portal adopts when reading/writing configuration parameters); thus its use in embedding the EW within DM-PLC's covert network is a logical next step. The open source library Snap7 \cite{Snap7} provides S7-Comm connectivity via its db\_read (retrieve data from a PLC, in the same way as GET) and db\_write (write data to a PLC, in the same way as PUT) functions. Listing~\ref{read} provides an example snippet of Python code, where Snap7 is being used to read one byte of data from PLC\_1, more specifically, Data Block 500 (a block of memory in the PLC), Byte 0. Listing~\ref{write} provides another example, however here Snap7 is being used to send one byte of data to PLC\_1, more specifically, the current value of variable ``poll'', to Data Block 501, Byte 0. The ability to read/write to and from data blocks over the network is common practice, and is employed by human machine interfaces (HMIs) that provide system operators with access to operational data and control capabilities~\cite{Green2021}.

\lstset{language=Python}
\lstset{frame=lines}
\lstset{caption={Snap7 Read from Data Block}}
\lstset{label={read}}
\lstset{basicstyle=\footnotesize}
\begin{lstlisting}
import snap7
client = snap7.client.Client()  
client.connect(172.21.1.101, 0, 1)
value = client.db_read(500, 0, 1)
client.disconnect()
print(value)
\end{lstlisting}
%\vspace*{-5mm}
\lstset{language=Python}
\lstset{frame=lines}
\lstset{caption={Snap7 Write to Data Block}}
\lstset{label={write}}
\lstset{basicstyle=\footnotesize}
\begin{lstlisting}
import snap7
client = snap7.client.Client()  
client.connect(172.21.1.101, 0, 1)
client.db_write(501, 0, poll)
client.disconnect()
\end{lstlisting}

As with PLC-PLC communications, setting up dummy data to send and request to/from each PLC is required, ensuring the communications library is performing as expected.

\subsubsection{Introduce Device Status Checkers}
\label{dsc}
\begin{figure}[h!]
    \includegraphics[width=1\linewidth]{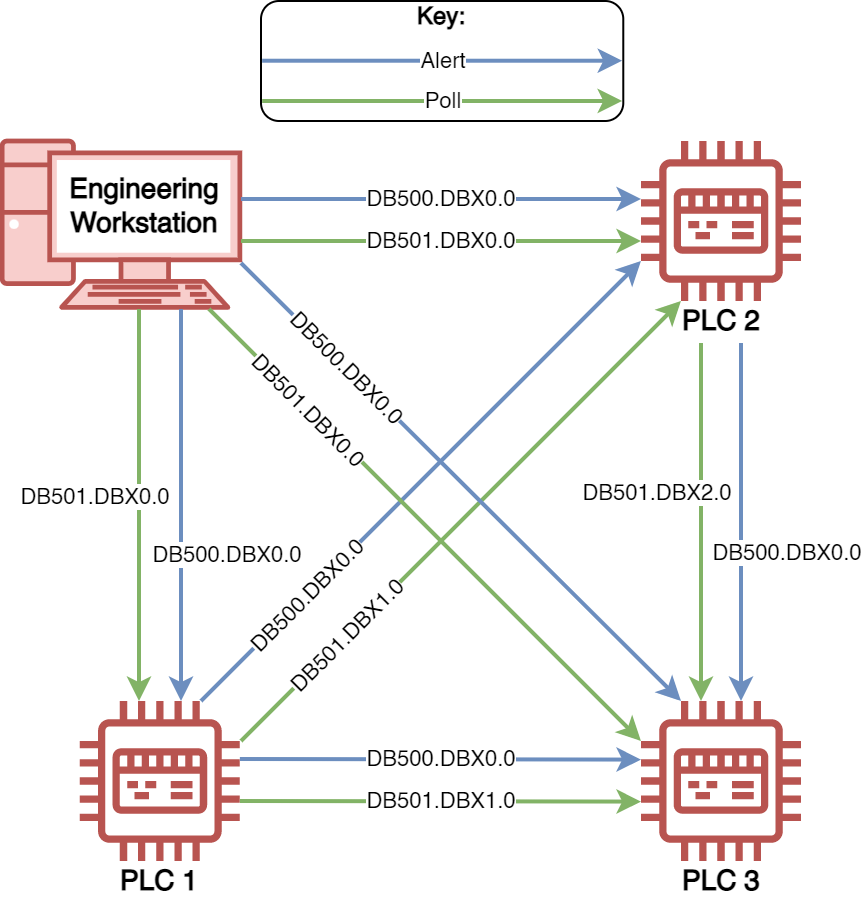}
    \vspace*{-5mm}
    \caption{DM-PLC's covert monitoring network example with 3 PLCs}
    \vspace*{-3mm}
    \label{fig:process}
\end{figure}

Having implemented a covert PLC-PLC and EW-PLC network, overlaying a resilient monitoring system to identify any attempt from the victim to recover their system is required. To support this discussion Figure~\ref{fig:process} has been included, acting as a reference point to demonstrate how the described approach can be scaled up as required, based on the size of the victim's environment.

Each device in the environment behaves in the same way, regardless of the communication library in use (i.e. PUT/GET on the PLCs, and Snap7 on the EW). Using PLC\_2 as a reference point for discussion, Figure~\ref{fig:process} depicts data entering and exiting this device. The EW will send its poll, a constantly changing boolean variable, to DB501.DBX0.0. This provides the following two key features:

\begin{itemize}
    \item The EW is able to ascertain if PLC\_2 is still online and operational. If PLC\_2 has been disconnected from the network, the EW will be unable to send this request, and will subsequently cease all polling actions across the entire environment (i.e., to PLC\_1, PLC\_2, and PLC\_3). Alternatively, as DB501.DBX0.0 has been created by the adversary, should it be removed, or have its access restricted in any way, the EW will again be unable to send its poll, and will cease all polling actions. 
    \item PLC\_2 is constantly monitoring DB501.DBX0.0 for state changes, should they cease (i.e. remain in a 0 or 1 state for more than a second), it will know that communications to the EW has either been removed by the victim, or the EW has been informed of another issue in the environment (e.g. it has observed an alert on PLC\_1, and has ceased all polling actions). PLC\_2 will then cease its own polling actions and raise an alert (set DB500.DBX0.0 to 1).
\end{itemize}

Continuing with PLC\_2 as an example, should it raise an alert and cease all polling actions, as it is no longer receiving a poll from the EW, the following two events will be realised:

\begin{itemize}
    \item PLC\_1 will see the alert as it is constantly monitoring address DB500.DBX0.0 in PLC\_2. In seeing this alert, it too will raise its own alert at the same address within its own memory (i.e. DB500.DBX0.0), and cease all of its polling actions.
    \item PLC\_3 will see PLC\_2 and PLC\_1 have both ceased their polling actions, and it too will raise an alert.
\end{itemize}

\begin{figure}[h!]
    \includegraphics[width=1\linewidth]{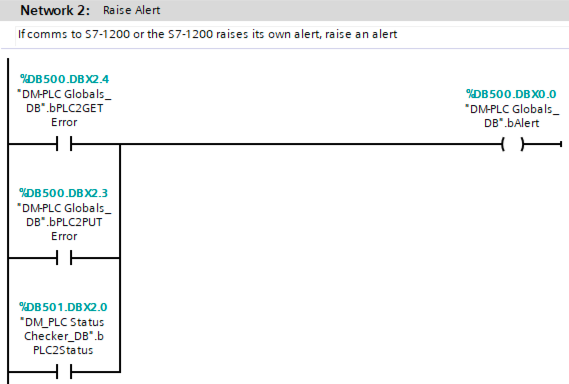}
    \vspace*{-4mm}
    \caption{PLC\_2 Monitoring and Alert Raising}
    \vspace*{-3mm}
    \label{fig:plc2monitoring}
\end{figure}

\begin{figure}[h!]
    \begin{center}
    \includegraphics[width=0.5\linewidth]{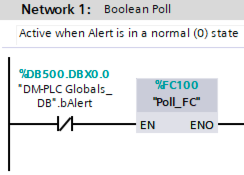}
    \end{center}
    \vspace*{-4mm}
    \caption{Disable Poll when Alert is Raised}
    \vspace*{-3mm}
    \label{fig:disablepoll}
\end{figure}

Figure~\ref{fig:plc2monitoring} shows how PLC\_1 monitors PLC\_2 and raises its own internal alert (i.e., should PLC\_1 fail to send/receive data from PLC\_2, or if PLC\_2 raises an alert). As previously discussed, PLC\_1 will then use its own alert as a trigger to cease all of its own polling actions (see Figure~\ref{fig:disablepoll} where polling is disabled).

\subsubsection{Introduce Code Supporting Operational Process Disruption}
\label{icsopd}
\begin{figure}[h!]
    \begin{center}
    \includegraphics[width=1\linewidth]{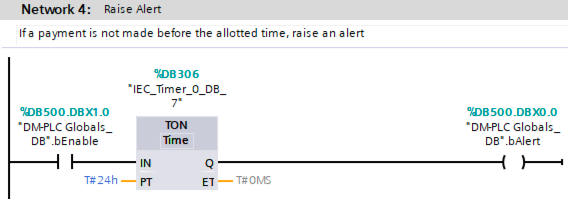}
    \end{center}
    \vspace*{-4mm}
    \caption{Payment Timeout}
    \vspace*{-3mm}
    \label{fig:timeout}
\end{figure}

Each PLC contains an alert bit, triggered through the detection of a victim attempting to recover their system, or when the payment timeout window is exceeded (see Figure~\ref{fig:timeout}). This alert is used to disable polling across the covert network (see Figure~\ref{fig:disablepoll}). However, its use within DM-PLC extends beyond this, acting as a trigger supporting operational process disruption.

\begin{figure}
    \begin{center}
    \includegraphics[width=0.6\linewidth]{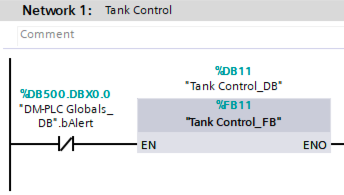}
    \end{center}
    \vspace*{-4mm}
    \caption{Disable Core Code Blocks}
    \vspace*{-3mm}
    \label{fig:disablecorecode}
\end{figure}

\begin{figure}
    \begin{center}
    \includegraphics[width=0.6\linewidth]{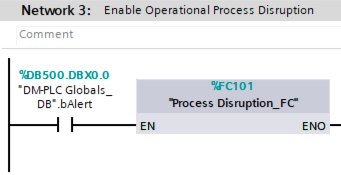}
    \end{center}
    \vspace*{-4mm}
    \caption{Enable Process Disruption Code Blocks}
    \vspace*{-3mm}
    \label{fig:enableopd}
\end{figure}

As a starting point, all legitimate operational code identified in Section~\ref{ccb} must first be disabled when the alert is raised. Figure~\ref{fig:disablecorecode} introduces the alert bit as a prerequisite to the execution of the Tank Control code block (see Figure~\ref{fig:ob1} before the alert prerequisite was introduced). It is also used to activate the operational process disruption code block (see Figure~\ref{fig:enableopd}).

\begin{figure}[h!]
    \begin{center}
    \includegraphics[width=0.75\linewidth]{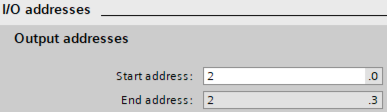}
    \end{center}
    \vspace*{-4mm}
    \caption{Digital Output Card Addressing}
    \vspace*{-3mm}
    \label{fig:docardio}
\end{figure}

In order to develop the operational process disruption code block, a review of each PLC's hardware profile must be undertaken. This provides a view of the I/O addressing mapped to physical output cards (i.e. Q2.0-2.3 in Figure~\ref{fig:docardio}).

\begin{figure}[h!]
    \begin{center}
    \includegraphics[width=1\linewidth]{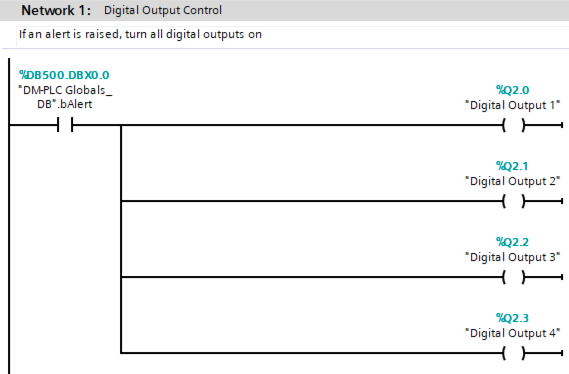}
    \end{center}
    \vspace*{-4mm}
    \caption{Turn All Digital Outputs On}
    \vspace*{-3mm}
    \label{fig:outputson}
\end{figure}

Once the output addresses have been confirmed, they can be put to use. For this PoC, all discovered digital outputs are set to an ON (1) state (see Figure~\ref{fig:outputson}). However, dependent upon the level of process comprehension an adversary is able to develop, this code block could become far more sophisticated, manipulating outputs to achieve maximum impact~\cite{Green2017}.

With regard to the EW, when DM-PLC is enabled all files and data on the system are encrypted as would be the case in a traditional encryption-based ransomware attack, including the original DM-PLC source code now that the process is running. 
The DM-PLC process itself is safe from tampering as any deviations from expected functionality will trigger the alert process.
If an alert is raised, the EW fails to poll any of the PLCs, or the allotted payment time is reached, the EW will simply shut down.

\subsubsection{Prevent Victims from Reversing all Changes}
\begin{figure}[h!]
    \begin{center}
    \includegraphics[width=0.85\linewidth]{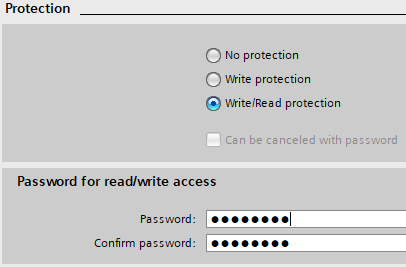}
    \end{center}
    \vspace*{-4mm}
    \caption{Setting the PLC Password}
    \vspace*{-3mm}
    \label{fig:plcpassword}
\end{figure}

\begin{figure}[h!]
    \begin{center}
    \includegraphics[width=0.85\linewidth]{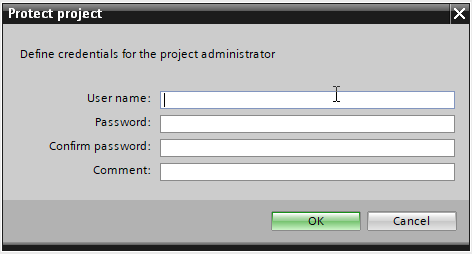}
    \end{center}
    \vspace*{-4mm}
    \caption{Setting the Project Password}
    \vspace*{-3mm}
    \label{fig:projectpassword}
\end{figure}

As discussed in Section~\ref{h}, OT vendors are increasing their cyber security capabilities. Turning unused capabilities against victims delivers not only critical functionality with minimal effort by the adversary, but is a lesson in making the most of what the adversary holds against the victim. Figures~\ref{fig:plcpassword} and~\ref{fig:projectpassword} provide a password for both the PLC and the TIA Portal project file. Without these two passwords, the victim is unable to read the malicious codebase from each PLC, reload their original trusted codebase to each PLC, or access the current modified version of their TIA Portal project file. 

From the perspective of the EW, victims are prevented from reverting any changes due to the deployment of encryption-based ransomware targeting all legitimate files and data.
As mentioned in Section \ref{icsopd}, the DM-PLC source code is subject to encryption-based ransomware and any attempts to tamper with the process will trigger the alert process.

\subsection{Arming DM-PLC}
\subsubsection{Arm DM-PLC}
\begin{figure}[h!]
    \begin{center}
    \includegraphics[width=0.5\linewidth]{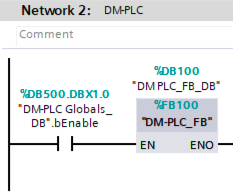}
    \end{center}
    \vspace*{-4mm}
    \caption{Enable DM-PLC}
    \vspace*{-3mm}
    \label{fig:enable}
\end{figure}

Once DM-PLC has been configured across all PLCs and the EW, it needs to be armed. This is executed by the adversary from the EW. For this PoC an ``enable'' boolean variable is introduced before the main DM-PLC code block on each PLC (see Figure~\ref{fig:enable}). The DM-PLC process running on the EW, providing it with access to the covert monitoring network (see Section~\ref{EWPLCComm}), is used to switch the enable variable to an ON (1) state in each PLC upon startup. This is performed using the Snap7 db\_write function from Listing~\ref{write}. Equally, should the victim make their payment on time, this same function is automatically used to switch each enable variable back to an OFF (0) state, deactivating DM-PLC across the entire estate. 

\section{Evaluation}
\label{evaluation}

While DM-PLC is a viable Cy-X technique for OT in concept and it was possible to implement it as expected, it cannot be considered practically viable without being tested in an established environment and having its outcomes evaluated. This section, therefore, describes the evaluation method, its results, and finally presents a discussion of those results and potential limitations of DM-PLC.

\subsection{Method}

DM-PLC was evaluated in the sterile conditions of an academically peer reviewed and industry validated OT testbed \cite{Green2020}, affording a sufficient level of realism without inducing any unnecessary risk to a live operational environment.
The evaluation utilised 3 PLCs and 1 EW, configured as depicted in Figure \ref{fig:process}.
The PLCs were Siemens ET200S (PLC 1), S7-300 (PLC 2), and S7-1200 (PLC 3), and the EW was running the Siemens TIA Portal V17 programming agent \cite{TIA}.
Overall, the evaluation took ~45 minutes to implement, with the majority of the time taken up by understanding the environment and configuring the first PLC.
Once the first PLC was configured, it was possible to duplicate its codebase to further PLCs rapidly with minimal edits, which reduces any additional configuration time and expedites the rest of the process.

The evaluation aimed to test three scenarios of DM-PLC:
\begin{enumerate}
    \item A PLC being removed from the network.
    \item The DM-PLC ransom timer expiring.
    \item The victim entering a code having `paid' their ransom.
\end{enumerate}

\subsection{Results}

The following presents the results of the 3 DM-PLC evaluation scenarios.

\begin{figure}
    \includegraphics[width=1\linewidth]{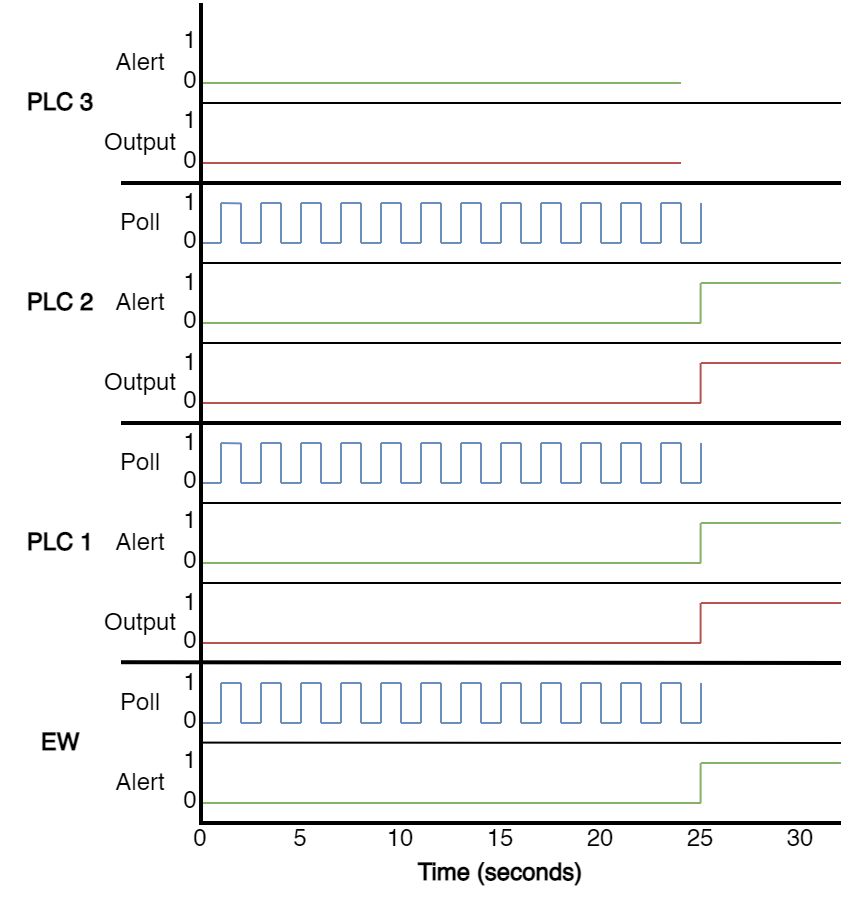}
    \vspace*{-5mm}
    \caption{PLC 3 being removed from the network}
    \vspace*{-5mm}
    \label{fig:plcremoved}
\end{figure}

\begin{figure}
    \includegraphics[width=1\linewidth]{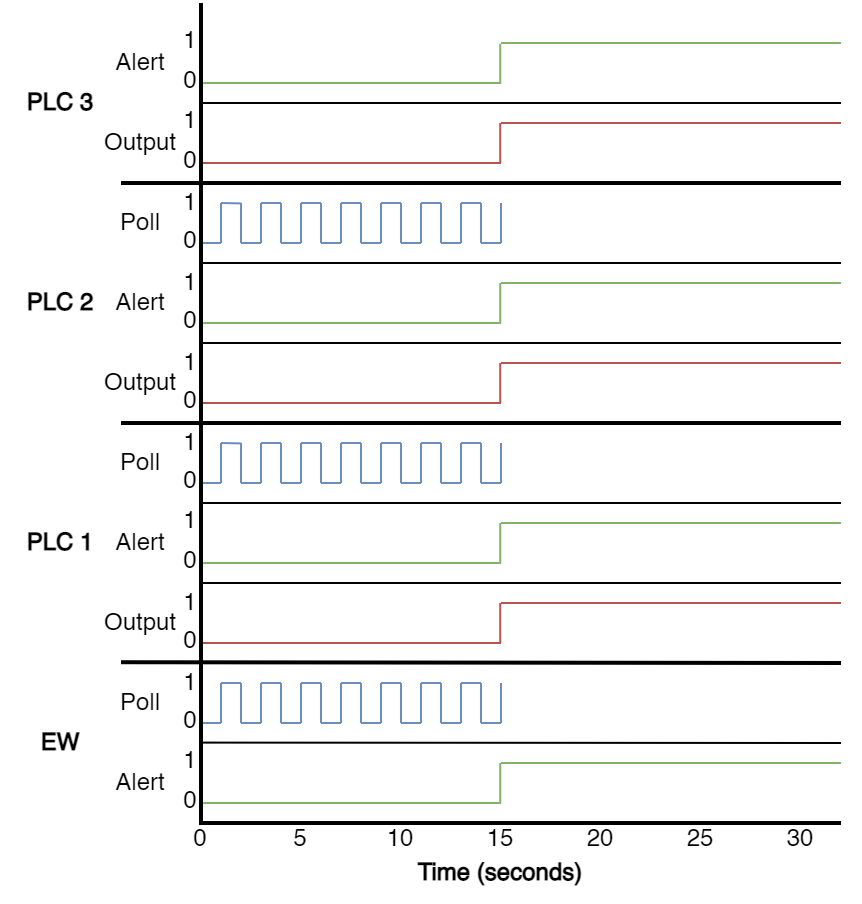}
    \vspace*{-5mm}
    \caption{3 PLCs and EW after time expiry}
    \vspace*{-8mm}
    \label{fig:plctimeout}
\end{figure}

\subsubsection{Scenario 1}
For the first scenario, DM-PLC was deployed and then armed, before removing the network cable from PLC 3.
As depicted in Figure \ref{fig:plcremoved}, once PLC 3's network cable was removed, it was no longer able to receive its polls and its alert bit became inaccessbile from experimental monitoring.
The final successful poll to PLC 3 was at 24 seconds, the poll at 25 seconds immediately failed from all devices, which all proceeded to set their alert bits to 1 and monitored sample outputs to ``ON''.

\subsubsection{Scenario 2}
The second scenario saw a ransom timer expiry being set to 15 seconds during configuration.
Figure \ref{fig:plctimeout} shows that at 15 seconds, the devices under evaluation interrupted their polling and immediately set their alert bits to 1 and monitored sample outputs to ``ON''.

\subsubsection{Scenario 3}
The final scenario tested in the evaluation saw the `victim' disarm DM-PLC, simulating a situation where they had paid their ransom and been given the deactivation key.
DM-PLC acted as expected and successfully disarmed, meaning that polling was interrupted and the alert bits were not set to 1 and no outputs were affected.

\subsection{Discussion and Limitations}

The three scenarios evaluated with DM-PLC were considered a success in that they provided the expected results.
The intended results for scenarios 1 and 2 were for all applicable PLCs and engineering workstations to raise an alert and cause operational impact by setting their outputs to ``ON'', which was successfully observed.
For scenario 3, the intended result was the successful disarmament of DM-PLC, which was also observed.
However, this does not mean that DM-PLC is without its limitations, particularly in its PoC form.
The rest of the section will discuss such limitations, and where appropriate, provide justification or potential mitigation to them.

\subsubsection{Dwell time}
Setting up DM-PLC requires gathering information to form an understanding of the environment and then configuration of the PLCs in that environment to deploy it, all of which takes time.
While this may seem like one of DM-PLC's major limitations, the median dwell time for an IT cyber extortion attack was reported to be 9 days by Mandiant \cite{mtrends}, which is likely to be sufficient time to prepare and deploy DM-PLC.

\subsubsection{Scalability}
The PoC for DM-PLC was conducted on 2 PLCs and its evaluation was conducted on just 3.
It is understandable that this may look to be a paltry number of devices to test upon when compared to large scale environments.
However, once the preparation is complete and the initial PLC is configured, copying and pasting code blocks and further minimal editing expedites the process considerably.
Furthermore, not every device within the OT environment has to be in the covert monitoring network; it may be sufficient to include only a subset of devices and simply apply the password protection to the rest, as once the consequences are enacted they will likely cascade.

Another issue of scalability is the concern whether DM-PLC is even possible on a large fleet of PLCs.
Depending on the PLC and number of communications processors, the number of simultaneous connections vary, but can potentially be as high as 92 \cite{scale}.
It is not inconceivable, therefore, that DM-PLC could be scaled to a significant subset, if not all PLCs, even in a large OT environment.

\subsubsection{Transitory network issues}
Due to the PoC setting its poll timer to be 1 second, it could be interpreted that DM-PLC is volatile and any transitory network issues could cause the Dead Man's switch to trigger.
However, the poll does not necessarily have to be 1 second, it could be set to a more forgiving interval.
Alternatively, a deadband could be introduced such that once polling stops, DM-PLC ensures there are multiple consecutive missed polls before triggering.

\subsubsection{Optimal damage}
In this PoC of DM-PLC the undesirable operational process disruption simply sets all outputs to an ``ON'' state, which may not be the most impactful method.
Should this be a real attack, the adversary could conduct further process comprehension to identify how to maximise DM-PLC's impact (e.g., turn a selection of pumps on, and open just one valve).

\subsubsection{Safe shutdowns}
One way a victim may recover from DM-PLC in its current PoC state is conducting a safe shutdown of their OT environment.
However, in the event of a real attack, minimal additional process comprehension could identify the signal to look out for and incorporate this as another trigger for DM-PLC's alert state.

\subsubsection{PoC weaknesses}
Two final potential limitations of DM-PLC are found in the deliberate simplicity of the PoC.
More specifically, the polling and enable/disable (arming/disarming) of DM-PLC are very simple binary functions.
If DM-PLC were to be used for a real attack, with resilience against recovery, these would ideally be more complex than just binary states.
\section{Mitigation}
\label{mitigation}
The implications of a successful DM-PLC deployment are significant. However, there are a number of steps organisations can take to better defend themselves. The following subsections discuss relevant mitigation techniques that could be employed at both device and network levels. These techniques not only act as preventative measures, but also provide alerting capabilities. Enhancing notifications to cyber security monitoring teams, and enabling the enactment of response and recovery practices forms a critical requirement. This is especially true where offensive techniques, such as DM-PLC, evolve over time, rendering preventative measure ineffective.

\subsection{Network}
Figure~\ref{fig:scenario} depicts the baseline attack scenario applied to the deployment of DM-PLC. While the area under consideration as part of this work excludes inter-connectivity between conventional IT environments and OT systems, it is important to note the criticality of their existence. Stepping up the DM-PLC kill-chain, appropriate management of remote access to EWs must be prioratised. Concepts within existing standards and guidelines, including Zones and Conduits~\cite{62443} can be used to better understand and classify operational assets and their place within a given infrastructure.

At a technical level, next generation security appliances, such as those provided by Check Point~\cite{CheckPoint}, offer enhanced industrial protocol recognition. This affords end-users with the ability to apply highly granular active rule-sets to control the flow of network traffic. For example, the Siemens S7Comm protocol used as part of the DM-PLC PoC, has a number of functions including upload, download, read, and write. Check Point is able to permit/deny traffic based on the function in use, its source, and destination. In addition to enhanced industrial protocol recognition, rule-sets can be applied to cover given time windows e.g., in/out of standard working hours where a system is under constant manual user control.

The notion of active network based controls within OT environments is often met with concerns over the prevention of legitimate network flows. Therefore, passive network traffic analysis may be more appropriate. Taking Claroty CTD~\cite{Claroty} as an example, like Check Point's next generation appliances, it too has enhanced industrial protocol recognition. Again, using Siemens S7Comm as an example, Claorty CTD is able to fully interpret data transactions within this protocol. However, unlike the Check Point active example, Claroty can operate without the need for rule-sets. Instead, through the application of machine learning techniques, Claroty builds a baseline of normal behavior by monitoring traffic flows over a given period of time, then raises alerts when it identifies deviations from this baseline. The covert network deployed as part of DM-PLC would be a prime example of such deviations.

\subsection{Engineering Workstation}
Although EWs exist in an industrial setting, it is important to create a clear distinction between them and neighbouring devices, such as Supervisory Control and Data Acquisiton (SCADA) systems, or PLCs, that provide a more critical role in the day to day management of operational processes. Therefore, conventional desktop-based system hardening techniques can be more aggressively applied. This can cover any number of categories, including the management of patches, services, applications, users, EDR, encryption, etc.

When considering OT-specific software operating on EWs, additional security features are becoming more prevalent. The most common of which is PLC project file encryption/password protection. As discussed in Section~\ref{implementation}, the DM-PLC PoC utilises Siemens TIA Portal project encryption as a method to restrict legitimate user access. In addition, software such as this requires a licence to function, which may reside on a vendor provided USB stick. If removed, the software will not operate. Removal of the licence while the system is not in use would further prevent unauthorised access. Finally, engineering software also includes a change log, which could be frequently reviewed for unexpected modifications.

\subsection{PLCs}
Due to the limited computational resources found in PLCs, embedded security features are often limited, however they are increasing. For example, in the DM-PLC PoC, the Siemens password protection feature is applied, protecting against unauthorised configuration uploads/downloads. Dependent upon the PLC in question, features such as this can be further expanded to limited connectivity between other devices, such as HMIs. Its is possible to restrict HMIs to read level permissions only, meaning that they cannot change setpoints, or alter the PLCs overarching configuration. Therefore, should a HMI be compromised, the impact it could cause is limited.

With some PLCs, applying protections to specific blocks of code is also possible. Such as know-how protection~\cite{SiemensKnowHow} in the Siemens eco system. This prevents an unauthorised person from accessing the underlying code within a code-block, meaning it cannot be read, or more importantly modified (a key requirement of DM-PLC).

While not directly operating on the PLC itself, PLC-focused forensic tooling is available. A recent open-source tool created by Microsoft~\cite{MSPLC} could be configured to run autonomously, actively connecting to PLCs and comparing their current configurations to project files. Any deviation in the live PLC configuration over the baseline project, such as those applied in DM-PLC, could be flagged for review by a security personnel.

\section{Conclusion}
\label{conclusion}

This work has introduced DM-PLC, a new approach to Cy-X targeted specifically against OT, which is in line with the modus operandi of an emerging adversary demographic in the area, viz. cyber criminals \cite{Forescout}.
As a counter to existing research that has focused solely on encryption-based ransomware, something that is trivially dealt with by existing response and recovery practices \cite{Staves2022}, DM-PLC follows existing trends in traditional Cy-X whereby adversaries are moving away from encryption-based ransomware tactics \cite{Navigator}.

DM-PLC creates a covert monitoring network of PLCs and EWs that constantly poll one another, and should any asset under adversary control deviate from the attack or the payment timeout window expire, an alert will be triggered akin to a Dead Man's switch, which will propagate throughout the covert monitoring network and turn all outputs to an ``ON'' state.
This is achieved by using only existing legitimate functionality, demonstrating that adversaries do not need to develop complex exploits against OT, therefore significantly reducing the potential cost of an OT cyber attack \cite{Derbyshire2021,Derbyshire2022}.

The approach presented in this paper should clearly serve as a warning to OT owners and operators that such an attack is feasible and needs special attention to defend against it.

Finally, further work in the area should focus on OT being secure by design such that existing functionality can not be weaponised \cite{Maesschalck2023,MSPLC}, rather than perpetuating existing practices of relying on re-appropriated IT security concepts.

%-------------------------------------------------------------------------------

\bibliographystyle{plain}
\bibliography{bibfile}

%%%%%%%%%%%%%%%%%%%%%%%%%%%%%%%%%%%%%%%%%%%%%%%%%%%%%%%%%%%%%%%%%%%%%%%%%%%%%%%%
\end{document}